\RequirePackage{ifpdf}
\ifpdf % We are running pdfTeX in pdf mode
\documentclass[pdftex]{sigma}
\else
\documentclass{sigma}
\fi

\newcommand {\mc} {\mathcal}
\newcommand {\la} {\lambda}
\newcommand {\al} {\alpha}

\begin{document}

\allowdisplaybreaks

\renewcommand{\PaperNumber}{075}

\FirstPageHeading

\ShortArticleName{Prolongation Loop Algebras for a Solitonic
System of Equations}

\ArticleName{Prolongation Loop Algebras\\ for a Solitonic System
of Equations}

\Author{Maria  A. AGROTIS}
\AuthorNameForHeading{M.A. Agrotis}

\Address{Department of Mathematics and Statistics, University of
Cyprus, Nicosia 1678, Cyprus}
\Email{\href{mailto:agrotis@ucy.ac.cy}{agrotis@ucy.ac.cy}}

\ArticleDates{Received September 13, 2006, in f\/inal form
November 01, 2006; Published online November 08, 2006}

\Abstract{We consider an integrable system of reduced
Maxwell--Bloch equations that describes the evolution of an
electromagnetic f\/ield in a two-level medium that is
inhomogeneously broadened. We prove that the relevant B\"acklund
transformation preserves the reality of the $n$-soliton potentials
and establish their pole structure with respect to the broadening
parameter. The natural phase space of the model is embedded in an
inf\/inite dimensional loop algebra. The dynamical equations of
the model are associated to an inf\/inite family of higher order
Hamiltonian systems that are in involution. We present the
Hamiltonian functions and the Poisson brackets between the
extended potentials.}


\Classification{37K10; 37N20; 35A30; 35Q60; 78A60}

\section{Introduction}

\looseness=-1 Integrable systems are closely related to the
inverse scattering method that serves as a means of integrating
the initial value problem, and also to inf\/inite dimensional Lie
algebras, else known as Kac--Moody Lie algebras or loop algebras.
Studying the underlying loop algebra reveals encoded properties of
the equations that are inherited from the integrable character of
the model. The Adler--Kostant--Symes (AKS) theorem gives the Lie
algebraic formulation of the dynamics of the system.

\looseness=1 In \cite{AKNS}, several nonlinear partial
dif\/ferential equations were realized as the compatibility
condition of two linear matrix systems of the form,
\begin{equation} \vec{v}_x=\begin{pmatrix}-i \la & q(x,t) \\ r(x,t)
& i \la \end{pmatrix} \vec{v}, \qquad \vec{v}_t=\begin{pmatrix} A
& B \\ C & -A
\end{pmatrix} \vec{v}, \label{ms} \end{equation} where $q$ and $r$
are the potentials satisfying a certain nonlinear evolution
equation and $A,B,C$ are functions of $x$ and $t$. The
Zakharov--Shabat dressing transformation \cite{Zak2} was then
employed to produce the soliton solutions of the system. The
matrices appearing on the right-hand-side of (\ref{ms}) are $2
\times 2,$ traceless matrices that can be viewed as elements of
the f\/inite dimensional Lie algebra $\mathfrak{sl}(2)$.
$\mathfrak{sl}(2)$ may be prolonged to an inf\/inite dimensional
Kac--Moody Lie algebra with the aid of the spectral parameter
$\la$, and lead to an inf\/inite number of systems in involution.
This is made exact in the context of the AKS theorem. Brief\/ly,
the theorem states that if one starts with a set of commuting
functions on a Lie algebra, then the corresponding Hamiltonian
systems are of course trivial. However, if we project those
functions to appropriate subalgebras then the resulting
Hamiltonian systems need not be trivial and continue to be in
involution.

In \cite{F1,F2,Adler1,Adler2,Kostant,Symes,FNR}, several
integrable equations have been studied in the context of the
inverse scattering technique and the AKS theorem. For example, the
Toda system and the Korteweg--de Vries equation, serving as
representatives of the ordinary and partial dif\/ferential
evolution equations respectively, have been associated to Lax pair
equations with one degree of freedom represented by the spectral
parameter. The phase space of the relevant model was extended to
include inf\/inite dimensional loop algebras. The loop algebra was
then decomposed into a vector space direct sum of subalgebras, and
with the aid of the trace functional it was identif\/ied with its
dual. In that way, the systems obtain the Kostant--Kirillov
symplectic structure, and an application of the AKS theorem
revealed an inf\/inite number of integrable Hamiltonian systems in
involution.

In this paper we consider a reduction of the Maxwell--Bloch
equations that models the optical pulse propagation of an electric
f\/ield through a two-level medium in the presence of an external
constant electric f\/ield. The optical resonance line of the
medium is inhomogeneously broadened. Following the terminology
used by McCall and Hahn \cite{MH} and Lamb \cite{Lamb2}, we refer
to inhomogeneous broadening as the phenomenon that occurs when the
atoms of the medium possess dif\/ferent resonant frequencies due
to microscopic interactions between them. In such a~case, the
induced electric dipole polarization is represented as a
continuum, and the resulting optical resonance line is
inhomogeneously broadened. In solids, such a broadening could be
caused by a distribution of static crystalline electric and
magnetic f\/ields and in gases by the distribution of Doppler
frequencies.

Since the late sixties and seventies with the papers
\cite{MH,Lamb2,Eilbeck}, among others, the Maxwell--Bloch
equations have undergone several treatments. Recently, various
reductions of the equations have been studied both analytically
and numerically. Lax pair operators, Darboux transformations and
soliton solutions were constructed and analyzed \cite{MC,E,QC,SU1}
and interesting applications in crystal acoustics have emerged
\cite{SU2,BU}.

Our scope in this paper is the study of the integrable structure
of a reduced Maxwell--Bloch system and the connections that arise
with Kac--Moody Lie algebras. In particular, we prove that the
B\"acklund transformation preserves the reality of the $n$-soliton
potentials $\forall\, n \in \mathbb{N},$ and establish their pole
structure with respect to the broadening parameter. The solitonic
phase space of the model is embedded in an inf\/inite dimensional
loop algebra and an application of the AKS theorem allows us to
view the system as a member of an inf\/inite family of systems in
involution. We present the higher order Hamiltonian functions and
f\/lows, as well as the Poisson brackets between the extended
potentials.

\section{Phase space}

\looseness=1 The optical equations we shall consider are the ones
presented in \cite{A2}. They model the propa\-gation of an
electric f\/ield in a two-level quantized medium, where the
optical resonance line of the medium has been inhomogeneously
broadened. The classical wave equation of Maxwell~\eqref{mb1} is
used for the evolution of a unidirectional electric f\/ield and is
coupled with the quantum mechanical Bloch equations
\eqref{mb2}--\eqref{mb4}, that describe the behavior of the
induced polarization f\/ield,
\begin{gather}
 \frac{\partial e}{\partial \zeta} + \frac{\partial e}{\partial
\tau}
= \left< \omega S_{\omega} \right>_g ,\label{mb1} \\
 \frac{\partial R_{\omega}}{\partial \tau}= (\beta -
\gamma e) S_{\omega}, \label{mb2} \\
 \frac{\partial S_{\omega}}{\partial \tau}=-(\beta - \gamma e)
R_{\omega}+
\frac{1}{2} \omega U_{\omega}, \label{mb3} \\
 \frac{\partial U_{\omega}}{\partial \tau}=- 2 \omega S_{\omega}
\label{mb4}.
\end{gather}
$\langle f\rangle_g=\int_{\infty}^{\infty} f(\omega) g(\omega)
d\omega,$ and denotes the weighted average of the function
$f(\omega)$ with respect to the distribution function,
\begin{equation*}
g(\omega)=\frac{\sigma}{\pi((\omega-\omega_0)^2+\sigma^2)}.
\end{equation*}
For a physical interpretation of the model see \cite{A2}. In this
paper we shall study the system \eqref{mb1}--\eqref{mb4} from a
Lie algebraic point of view.

The system is completely integrable and admits a Lax pair
representation. We def\/ine the dif\/ferential operators $L$ and
$A$ whose commutativity, $[L,A]:=LA-AL=0$, is equivalent to
equations (\ref{mb1})--(\ref{mb4})
\begin{gather*}
A= -\partial_{\tau} +Q^{(0)}, \qquad L= \partial_{\zeta} + Q^{(1)}
, \nonumber
\end{gather*}
where,
\begin{gather*}
Q^{(0)}= \lambda (h_0 \mathcal{H} + f_0 \mathcal{F}) + e_0 \mathcal{E}, \\
Q^{(1)}= \lambda (h_0 \mathcal{H} + f_0 \mathcal{F}) + e_0
\mathcal{E} + \int_{-\infty}^{\infty} \frac{1}{(\omega^2 -
\lambda^2)} [\lambda ( h_1 \mathcal{H} + f_1 \mathcal{F} ) + e_1
\mathcal{E}] d \omega,
\end{gather*}
and
\begin{alignat}{3}
& h_0=\tfrac{1}{2}, &&   h_1= - \tfrac{1}{2} \gamma \omega
g(\omega)
R_{\omega},&\nonumber \\
& f_0=0, && f_1=- \tfrac{1}{2} \gamma \omega g(\omega)
S_{\omega}, & \label{coeffs}\\
& e_0= - \tfrac{1}{2} (\beta - \gamma  e),\qquad & &
e_1=\tfrac{1}{4} \gamma \omega^2 g(\omega) U_{\omega}.&\nonumber
\end{alignat}
$\mc{H}$, $\mc{F}$ and $\mc{E}$ form a basis of the semi-simple
Lie algebra $\mathfrak{su}(2)$, and are given as follows,
\begin{equation*} \mc{H}=\begin{pmatrix} i & 0 \\ 0 & -i \end{pmatrix}, \qquad \mc{F}=
\begin{pmatrix}0 & 1 \\ -1 & 0\end{pmatrix},\qquad  \mc{E}=\begin{pmatrix} 0 & i \\ i & 0
\end{pmatrix} .\end{equation*} We call $h_{j}$, $f_{j}$, $e_{j}$ for $j=0,1$, the
potentials and $Q^{(0)}$, $Q^{(1)}$ loop elements because they can
be considered as elements of an inf\/inite dimensional loop
algebra that we will def\/ine in Section~\ref{embedding}. We note
that the potentials depend on the solutions $e$, $R_{\omega}$,
$S_{\omega}$, $U_{\omega}$ of the inhomogeneously broadened
reduced Maxwell--Bloch (ib-rMB) equations
(\ref{mb1})--(\ref{mb4}). The commutation of the dif\/ferential
operators $L$ and $A$ gives rise to the following Lax pair
equation
\begin{equation} \frac{\partial Q^{(0)}} {\partial \zeta} + \frac{\partial
Q^{(1)}} {\partial \tau}=[Q^{(0)},Q^{(1)}], \label{lpe}
\end{equation} which is equivalent to the ib-rMB system.

The Lax pair can be used to construct a B\"acklund transformation
(BT) that iteratively produces the soliton solutions of equations
(\ref{mb1})--(\ref{mb4}).

We consider the spectral problem,
\begin{equation*}
\partial_{\tau} \Psi=Q^{(0)} \Psi, \qquad
\partial_{\zeta} \Psi=-Q^{(1)} \Psi,\end{equation*}
and aim to f\/ind a new eigenfunction $\Psi$ and the corresponding
new loop element $Q^{(1)}$ that satisfy the spectral problem. The
loop elements are functions of $h_j$, $f_j$, $e_j,$ for $j=0,1$
and will, in turn, give rise to the new solutions of the ib-rMB
system via expressions (\ref{coeffs}). This transformation theory
leads to an analogue of superposition formulas that allows one to
construct multi-soliton solutions starting from single solitons by
algebraic means \cite{B,L,LA,N,QC}. We brief\/ly describe the
procedure and quote the relevant theorem from reference \cite{A2}.

One begins with a constant solution to equations
(\ref{mb1})--(\ref{mb4}), which in turn determines
potentials~(\ref{coeffs}) and the corresponding loop element, call
it $Q_0$. We then f\/ind a simultaneous, fundamental solution
$\Psi_{1}$ to the Lax pair system $L\Psi=0, A\Psi=0$ and def\/ine
$\vec{\Phi}_1=\begin{pmatrix}
           \phi_1^1 \\
           \phi_2^1
           \end{pmatrix} := \Psi_1(\lambda=\nu_1) \vec{c_1},$
where $\vec{c_1}=\begin{pmatrix}
           c_1^1 \\
           i c_2^1
            \end{pmatrix}$ is a constant vector with $c_1^1,c_2^1 \in \mathbb{R}$,
and the matrix $N_1$ as,
\begin{equation*}
N_1=\begin{pmatrix}
    \phi_1^1 & - \overline{\phi_2^1} \\
     \phi_2^1 & \overline{\phi_1^1}
    \end{pmatrix}.
\end{equation*}
The BT matrix function is constructed as: \[ G(\nu_1,
\vec{c_1};\lambda)= N_1
   \begin{pmatrix} \lambda-\nu_1 & 0 \\
                   0 &  \lambda-\overline{\nu_1}
   \end{pmatrix} N_1^{-1}.\] Applying $G$ to $\Psi_1$ yields a new fundamental solution:
$ \Psi_2(\nu_1, \vec{c_1};\lambda)= G(\nu_1, \vec{c_1};\lambda)
\Psi_1(\nu_1, \vec{c_1};\lambda),$ and the procedure is iterated.
The formula for the loop element after $n$ iterations of the BT,
call it $Q_n$, in terms of the previous one $Q_{n-1}$ and the
matrix $N_n$ is the context of the next theorem.
\begin{theorem} \label{newloop}
\begin{gather}
 Q_n(\lambda)=\lambda h_0^{n-1} \mathcal{H} +
   m_n h_0^{n-1} [\mathcal{H}, N_n\mathcal{H} N_n^{-1}]
   + e_0^{n-1} \mathcal{E}  +  \int_{-\infty}^{\infty} \frac{1}{(\omega^2 - \lambda^2)}
   \frac{1}{(\omega^2 + m_n^2)}  \nonumber \\
\phantom{Q_n(\lambda)=}{}\times \left\{ \lambda \left[ \omega^2
(h_1^{n-1} \mathcal{H}+
   f_1^{n-1} \mathcal{F}) -  (m_n)^2
   (N_n \mathcal{H} N_n^{-1}) (h_1^{n-1} \mathcal{H}
   + f_1^{n-1} \mathcal{F}) (N_n \mathcal{H} N_n^{-1})
  \right. \right. \nonumber \\
 \left.\phantom{Q_n(\lambda)=}{} +  m_n e_1^{n-1}
   [\mathcal{E}, N_n \mathcal{H} N_n^{-1}] \right]
   + m_n \omega^2 \left( h_1^{n-1} [\mathcal{H}, N_n \mathcal{H} N_n^{-1}]+
   f_1^{n-1} [\mathcal{F}, N_n \mathcal{H} N_n^{-1}] \right) \nonumber \\
 \left.\phantom{Q_n(\lambda)=}{}+\omega^2 e_1^{n-1}\mathcal{E} - (m_n)^2 e_1^{n-1}
   (N_n \mathcal{H} N_n^{-1}) \mathcal{E} (N_n \mathcal{H} N_n^{-1}) \right\}
   d\omega .\label{newloopeq}
\end{gather}
\end{theorem}

We have taken the specif\/ic value of the spectral parameter to be
purely imaginary, $\nu_n=i m_n \in i\mathbb{R}.$ The general form
of the $n$-soliton loop is given by,
\begin{equation} \label{qni}
Q_n(\lambda)=\la(h_0 \mc{H} +f_0 \mc{F}) +e_0^n \mc{E} +
\int_{-\infty}^{\infty} \frac{1}{(\omega^2-\lambda^2)}[ \lambda
(h_1^n \mc{H}+f_1^n \mc{F})+ e_1^n\mc{E}] d\omega,
\end{equation} where the upper index of
$e_0^n$, $h_1^n$, $f_1^n$, $e_1^n$ and the lower index of $Q_n$,
$\Psi_n$, $\vec{\Phi}_n$, $N_n$ indicate the level of the
B\"acklund transform. We note that $h_0$ and $f_0$ are constant
functions of space and time and are invariants of the level of the
BT. To ensure the reality of the potentials and consequently of
the solutions $e$, $R_{\omega}$, $S_{\omega}$, $U_{\omega},$ we
compare (\ref{newloopeq}) and (\ref{qni}) and impose the following
conditions:
\begin{gather} \label{cond1a}
a) \ \ [\mathcal{H},N_n \mathcal{H} N^{-1}_n], [\mc{F},N_n \mc{H}
N^{-1}_n],
(N_n \mathcal{H} N^{-1}_n)\mc{E}(N_n \mathcal{H} N^{-1}_n)\in {\rm span}\,\{\mathcal{E}\}, \\
b) \ \ (N_n \mathcal{H} N^{-1}_n)\mc{H}(N_n \mathcal{H} N^{-1}_n),
(N_n \mathcal{H} N^{-1}_n)\mc{F}(N_n \mathcal{H} N^{-1}_n),
[\mc{E},N_n \mc{H} N^{-1}_n] \in {\rm
span}\,\{\mc{H,\mc{F}}\}.\label{cond1b}
\end{gather}
By def\/inition
\begin{equation*}
N_n=\mbox{Re}(\phi_1^n) \mc{I}+\mbox{Im}(\phi_1^n) \mc{H} -
\mbox{Re} (\phi_2^n) \mc{F} + \mbox{Im} (\phi_2^n) \mc{E},
\end{equation*} where $\mc{I}$ is the 2$\times$2 identity matrix.
One can f\/ind that conditions (\ref{cond1a})--(\ref{cond1b}) are
satisf\/ied if and only if
\begin{equation}
\mbox{Im}(\phi_1^n) \mbox{Im}(\phi_2^n) + \mbox{Re} (\phi_1^n)
\mbox{Re} (\phi_2^n) = 0 .\label{cond3}
\end{equation}
We can construct $\phi_1^n,\phi_2^n$ such that one of the
following two cases holds: a) $\phi_1^n \in \mathbb{R}$, $\phi_2^n
\in i\mathbb{R}$, or b)~$\phi_1^n \in i \mathbb{R}$, $\phi_2^n \in
\mathbb{R}$. This is the context of the following proposition.
\begin{proposition}
At any given level of the B\"acklund transformation the reality of
the $n$-soliton potentials can be secured by an appropriate choice
of the transformation data.
\end{proposition}
\begin{proof}
A constant set of solutions to equations (\ref{mb1})--(\ref{mb4})
is given by $e=\frac{\beta}{\gamma}$, $S_{\omega}=0$,
$U_{\omega}=0$ and $R_{\omega}=R^{\rm init}$ a nonzero constant.
The corresponding potentials become $ h_0=\frac{1}{2}$,
$h_1=-\frac{1}{2}\gamma \omega g(\omega)R^{\rm init}$, $f_0= f_1=
e_0= e_1=0.$ Following the procedure of the B\"acklund transform
we f\/ind, $ \phi_1^1=\mbox{Re}(c_1^1) e^{x_1}+i \mbox{Im}(c_1^1)
e^{x_1}$, $\phi_2^1=\mbox{Re}(c_2^1) e^{-x_1}+i \mbox{Im}(c_2^1)
e^{-x_1},$ where $x_1=2((v_1+m_1h_0) \zeta - m_1 h_0 \tau) \in
\mathbb{R}$, and $v_1$ is an expression independent of $\omega,
\la,\zeta$ and $\tau$. We choose $c_1^1 \in \mathbb{R}$ and $c_2^1
\in i\mathbb{R}$ so that $\phi_1^1 \in \mathbb{R}$, $\phi_2^1 \in
i\mathbb{R}$ and condition (\ref{cond3}) holds for $n=1$. The
proof that the condition is satisf\/ied at any given level of the
BT lies on the following:

$N_{2n-1} \in \mbox{span} \{\mc{I},\mc{E}\},$ and $N_{2n} \in
\mbox{span} \{\mc{H},\mc{F}\}.$ This yields $G_n(\lambda) \in
\mbox{span}\{\lambda I, \mc{H},\mc{F}\}$, $\forall\, n \in
\mathbb{N}.$ Using the def\/inition $\Psi_{n}=G_n(\lambda=im_n)
\Psi_{n-1}(\la=im_n)$ we deduce that $\Psi_{2n-1} \in
\left\{\begin{pmatrix}
\mathbb{R} & i\mathbb{R} \\
i\mathbb{R} &  \mathbb{R}
\end{pmatrix}\right\},$ and $\Psi_{2n} \in
\left\{\begin{pmatrix}
i\mathbb{R} & \mathbb{R} \\
\mathbb{R} & i \mathbb{R}
\end{pmatrix}\right\}$. Therefore, if we choose  $\vec{c}_n
\in \left\{\begin{pmatrix}  \mathbb{R} \\
                 i\mathbb{R}\end{pmatrix}\right\}$, and use
                 $\vec{\Phi}_n=\vec{\Psi}_n \vec{c}_n,$ we
                 obtain that
$\vec{\Phi}_{2n-1} \in \left\{\begin{pmatrix}  \mathbb{R} \\
               i \mathbb{R}\end{pmatrix}\right\},$ and $\vec{\Phi}_{2n}
               \in \left\{\begin{pmatrix} i\mathbb{R} \\
               \mathbb{R}\end{pmatrix}\right\}$ as desired.
\end{proof}

We aim to give an integral-free representation of $Q_n(\lambda)$
so that its $\lambda$-structure becomes apparent. We begin with
$Q_1(\lambda)$ and then generalize the construction for
$Q_n(\lambda)$. The general form of $Q_1(\lambda)$ is the
following,
\begin{equation*}
 Q_1(\lambda)=\la(h_0 \mc{H} +f_0 \mc{F}) +e_0^1 \mc{E} +
\int_{-\infty}^{\infty} \frac{1}{(\omega^2-\lambda^2)}\big[
\lambda (h_1^1 \mc{H}+f_1^1 \mc{F})+ e_1^1\mc{E}\big] d\omega.
\end{equation*} The one soliton potentials are given as,
\begin{gather*}
e_0^1=2m_1 h_0 \,\mbox{sech}(x_1),\qquad
e_1^1=\frac{2 m_1 \omega^2 h_1} {(\omega^2+m_1^2)} \,\mbox{sech}(x_1), \\
f_1^1=\frac{2 m_1^2 h_1} {(\omega^2+m_1^2)} \,\mbox{sech}(x_1)
\tanh(x_1), \qquad h_1^1=\frac{h_1} {(\omega^2+m_1^2)}
\big[\omega^2+m_1^2-2 m_1^2\, \mbox{sech}^2(x_1)\big] ,
\end{gather*}
where $x_1=2((v_1+m_1 h_0)\zeta-m_1 h_0 \tau)+ \ln(c_1^1/c_2^1),$
$h_1=-\frac{1}{2}\gamma \omega g(\omega)R^{\rm init}$ and
$v_1=-\frac{1}{2} \frac{\gamma R^{init} m_1
\omega_0}{((m_1+\sigma)^2+\omega_0^2)}.$ We write the last three
in the following more convenient form,
\begin{gather}
e_1^1=\frac{\beta_1 \omega^3} {(\omega^2+m_1^2)((\omega-\omega_0)^2+\sigma^2)}, \label{p1}\\
f_1^1=\frac{\beta_2 \omega} {(\omega^2+m_1^2)((\omega-\omega_0)^2+\sigma^2)}, \label{p2} \\
h_1^1=\frac{\beta_3\omega^3+\beta_4\omega}{(\omega^2+m_1^2)((\omega-\omega_0)^2+\sigma^2)},
\label{p3}
\end{gather}
where $\beta_1$, $\beta_2$, $\beta_3$, $\beta_4$ are functions of
$\zeta$ and $\tau$, but do not depend on $\omega$ or $\lambda$. We
will use contour integration and Cauchy's integral formula to
compute~$Q_1$. We def\/ine the complex-valued function,
\begin{equation} \nonumber
h(z)=\frac{z}{(z^2-\lambda^2)(z^2+m_1^2)((z-\omega_0)^2+\sigma^2)}\big\{
\lambda [(\beta_3 z^2+\beta_4) \mc{H}+\beta_2 \mc{F}]+ \beta_1 z^2
\mc{E} \big\}, \end{equation} which has three poles in the upper
half complex plane at $z_1=\lambda$, $z_2=im_1,$ and
$z_3=\omega_0+i \sigma$. We integrate around a simple contour that
consists of a semicircle of radius $R$ in the upper-half complex
plane, call it $C_R$, and the segment on the real axis from $-R$
to $R$. We choose $R$ big enough to include all the poles of
$h(z)$ that appear in the upper-half complex plane. It is not hard
to see that $ \lim\limits_{R \mapsto \infty} \int_{C_R}h(z) dz=0,$
and thus using the Cauchy integral formula we obtain,
\begin{equation} Q_1=\la(h_0 \mc{H}+f_0 \mc{F})+ e_0^1 \mc{E} + 2 \pi i
\sum_{k=1}^3 \mbox{Res}_{z=z_k} h(z), \nonumber\end{equation}
which can be written as,
\begin{gather} \nonumber Q_1(\lambda) =  \la(h_0 \mc{H}+f_0 \mc{F})+
e_0^1 \mc{E} + \frac{1}{(\lambda^2+m_1^2)}[\lambda(\delta_1
\mc{H}+\delta_2 \mc{F})+\delta_3 \mc{E}] \\
  \phantom{Q_1(\lambda) =}{} + \frac{
\lambda(\tilde{\delta_1} \mc{H}+ \tilde{\delta_2}
\mc{F})+\tilde{\delta_3} \mc{E}+\lambda^3
(\tilde{\tilde{\delta_1}} \mc{H}+ \tilde{\tilde{\delta_2}}
\mc{F})+\la^2 \tilde{\tilde{\delta_3}}
\mc{E}}{((\omega_0^2-\sigma^2- \lambda^2)^2+4 \omega_0^2
\sigma^2)}, \label{q1} \end{gather} where $\delta_j$,
$\tilde{\delta}_j$, $\tilde{\tilde{\delta}}_j$, $j=1,2,3$ are
independent of $\la$ and $\omega$. Let $S_1=\{\delta_1, \delta_2,
\delta_3\}$, $S_2=\{\tilde{\delta_1} \tilde{\delta_2},
\tilde{\delta_3}\}$ and $S_3=\{\tilde{\tilde{\delta_1}}
\tilde{\tilde{\delta_2}}, \tilde{\tilde{\delta_3}}\}$. The
elements of each set $S_1$, $S_2$ or $S_3$ are functions of the
one-soliton potentials and they obey the ib-rMB equations.

We will use induction to obtain the general form of the
$n$-soliton potentials $h_1^n$, $f_1^n$, $e_1^n$.
\begin{proposition}
The pole structure with respect to the broadening parameter
$\omega$ of the general $n$-soliton potentials of the ib-rMB
system is the following:
\begin{gather*}
h_1^n=\frac{\omega \sum\limits_{k=0}^{n}\gamma_k
\omega^{2k}}{((\omega-\omega_0)^2+\sigma^2)
\prod\limits_{k=1}^{n}(\omega^2+m_k^2)}, \qquad f_1^n=\frac{\omega
\sum\limits_{k=0}^{n}\delta_k
\omega^{2k}}{((\omega-\omega_0)^2+\sigma^2)
\prod\limits_{k=1}^{n}(\omega^2+m_k^2)}, \\
e_1^n=\frac{\omega \sum\limits_{k=0}^{n}\epsilon_k
\omega^{2k}}{((\omega-\omega_0)^2+\sigma^2)
\prod\limits_{k=1}^{n}(\omega^2+m_k^2)}.
\end{gather*}
\end{proposition}
\noindent We note that $\gamma_k,\delta_k,\epsilon_k,\:
k=1,\ldots,n,$ are analytic functions of $\zeta$ and $\tau,$ and
do not depend on $\omega.$
\begin{proof}
\noindent The proposition holds for $n=1$ as can be seen by
(\ref{p1})--(\ref{p3}). We assume that the proposition holds for
$n-1$ and show that it holds for $n$. Using the induction
hypothesis in Theorem \ref{newloop}
 and the fact that $N_{k} \mc{H}N_{k}^{-1} \in \mbox{span}\{\mc{H}, \mc{F}\}$
 for any $k \in \mathbb{N}$ we get,
\begin{gather*}
 Q_n(\lambda) = \la(h_0 \mc{H}+f_0 \mc{F})+ e_0^n \mc{E} +
\int_{-\infty}^{\infty} \frac{\omega}{(\omega^2 - \lambda^2)
   ((\omega-\omega_0)^2+\sigma^2) \prod\limits_{k=1}^n(\omega^2 + m_k^2)}  \nonumber \\
\phantom{Q_n(\lambda) =}{}\times \left\{ \lambda \left[
\sum_{k=0}^n \gamma_k \omega^{2k} \mathcal{H}+ \sum_{k=0}^n
\delta_k \omega^{2k} \mathcal{F}\right] + \sum_{k=0}^n \epsilon_k
\omega^{2k} \mathcal{E}  \right\}
   d\omega. \label{qn}
\end{gather*}
By def\/inition
\begin{equation}
 Q_n(\lambda)=\la(h_0 \mc{H}+f_0 \mc{F})+ e_0^n \mc{E} +
\int_{-\infty}^{\infty} \frac{1}{(\omega^2-\lambda^2)}[ \lambda
(h_1^n \mc{H}+f_1^n \mc{F})+ e_1^n\mc{E}] d\omega. \label{iqn}
\end{equation}
Therefore, by equating the last two expressions of $Q_n(\la)$ we
prove the proposition. We note that some of the coef\/f\/icients
$\gamma_k$, $\delta_k$, $\epsilon_k$ may be equal to zero. For
example, for $n=1$, $\delta_1=0$ and $\epsilon_0=0$.
\end{proof}
The form of the $n$-soliton potentials can be used to identify the
$\la$-structure of a general $n$-soliton loop element $Q_n(\la)$.
\begin{proposition}
\begin{gather}
 Q_n(\lambda) =  \la(h_0 \mc{H}+f_0 \mc{F})+ e_0^n \mc{E}
+ \frac{\lambda(\tilde{\delta_1} \mc{H}+ \tilde{\delta_2}
\mc{F})+\tilde{\delta_3} \mc{E}+\lambda^3
(\tilde{\tilde{\delta_1}} \mc{H}+ \tilde{\tilde{\delta_2}}
\mc{F})+\la^2 \tilde{\tilde{\delta_3}} \mc{E}
}{((\omega_0^2-\sigma^2-\lambda^2)^2+4 \omega_0^2 \sigma^2)}
\nonumber \\
 \phantom{Q_n(\lambda) =}{} +
\sum_{k=1}^{n}\frac{1}{(\lambda^2+m_k^2)}[\lambda(\delta_1^k
\mc{H}+\delta_2^k \mc{F})+\delta_3^k \mc{E}] . \label{ifqn}
\end{gather}
\end{proposition}
\noindent The proof follows the same idea that was used to derive
the integral-free form (\ref{q1}) of $Q_1(\lambda)$ and is
omitted.

Let $S_k=\{\delta_1^k, \delta_2^k, \delta_3^k\}$, $k=1,\ldots,n$,
$S_{n+1}=\{\tilde{\delta_1}, \tilde{\delta_2}, \tilde{\delta_3}\}$
and $S_{n+2}=\{\tilde{\tilde{\delta_1}}, \tilde{\tilde{\delta_2}},
\tilde{\tilde{\delta_3}}\}.$ The elements of each set $S_k$, $
k=1,\ldots,n+2$ are functions of the $m$-soliton potentials for
$m=1,\ldots,k$ and they satisfy the ib-rMB equations.

\section{Embedding} \label{embedding}

We call $\alpha_k=im_k$, $k=1,\ldots,n$, $\alpha_{n+1}=\omega_0 -
i \sigma$ and $\alpha_{n+2}=\omega_0+i\sigma$. Then $Q_n(\la)$ of
(\ref{ifqn}) can be rewritten as
\begin{equation} Q_n(\la)=\la(h_0 \mc{H} + f_0 \mc{F}) + e_0 \mc{E} +
\sum_{k=1}^{n+2} \frac{X_{k}^-}{\la-\al_k} +
\frac{X_{k}^+}{\la+\al_k}, \nonumber
\end{equation} where
$X_{k}^-$, $X_{k}^+$ are linear combinations of the basis elements
of the Lie algebra $\mathfrak{su}(2)$, $\{\mc{H},\mc{F},\mc{E}\}$.

We def\/ine the following inf\/inite dimensional Lie algebra
$Q^{\rm ext}$ that extends loops of the form~(\ref{ifqn}),
\begin{equation}  Q^{\rm ext} = \left\{ X: \: X=\sum_{j=0}^{\infty}
\sum_{k=1}^{n+2} \frac{1}{(\lambda^2-\al_k^2)^j}[\lambda(h_j^k
\mc{H}+f_j^k \mc{F})+e_j^k \mc{E}] \right\}.
\label{qext}\end{equation} We note that the coef\/f\/icients
$h_j^k$, $f_j^k$, $e_j^k$ of (\ref{qext}) are not the same as
those of (\ref{iqn}) where the upper index denotes the level of
the BT. In (\ref{qext}) $X$ is an element that extends the natural
solitonic phase space of the ib-rMB equations and the upper index
indicates the relevant pole $\al_k$, whereas the lower index
indicates the order of the pole $\al_k^{2}$.

We embed this Lie algebra into a larger one,
\begin{equation}  Q = \left\{ X : \:X= \sum_{j=1}^{\infty} \left(X_{j-1}
\la^{j-1} + \sum_{k=1}^{2(n+2)} \frac{X_j^k}{(\lambda-\al_k)^j}
\right)\right\}, \label{q}\end{equation} where we have set
$\al_{k+n+2}=-\al_k,$ for  $k=1,\ldots, n+2$. We note that for $Q$
to be a Lie algebra a f\/initeness condition needs to imposed.
Namely, $X_{j}=0$ and $X_j^k=0$, $\forall \, k=1,\ldots,2(n+2)$
and $j \geq j_0$ for some $j_0 \in \mathbb{N}.$

\section[Application of the Adler-Kostant-Symes theorem]{Application of the Adler--Kostant--Symes theorem}
\label{aks}

Following the ideas in the theorem of Adler, Kostant and Symes
\cite{Adler1,Adler2,Kostant,Symes}, we decompose the inf\/inite
dimensional loop algebra $Q$ into a direct sum of two subalgebras
$a$ and $b$, and def\/ine an ad-invariant, non-degenerate inner
product on $Q$. The perpendicular complements $a^{\perp}$,
$b^{\perp}$ with respect to the inner product serve as another
direct sum decomposition of $Q$. Using the Riesz representation
theorem and the non-degenerate inner product one may def\/ine an
isomorphism between $a^{\perp}$ and $b^*,$ the dual of the Lie
subalgebra $b$. The canonical Lie--Poisson bracket that exists on
$b^*$ is then represented on $a^{\perp}.$ If necessary, one may
translate $a^{\perp}$ by any element $\beta \in b^{\perp}$ that
satisf\/ies $(\beta,[x,y])=0$ $\forall \, x,y  \in a$, so that the
translated space $\beta+a^{\perp}$ includes the natural phase
space of the relevant system \cite{A}.

We write $Q$ as the vector space direct sum of the Lie subalgebras
\begin{gather}
a = \left\{X \in Q : \  X= \sum_{j=1}^{\infty} \sum_{k=1}^{2(n+2)}
\frac{X_j^k}{(\lambda-\al_k)^j} \right\}, \qquad b = \left\{X \in
Q : \  X= \sum_{j=0}^{\infty} X_{j} \la^{j} \right\} \nonumber.
\end{gather} A non-degenerate inner product is def\/ined on $Q$
using the trace map. Namely,
\begin{gather}
Q \times Q \longrightarrow \mathbb{C},\qquad (X,Y) \mapsto
\mbox{Tr}(XY)_0, \nonumber
\end{gather} where $(XY)_0$ denotes the matrix coef\/f\/icient of $\la^0$ in
the product $XY$. We note that the inner product is ad-invariant.
That is, $(X,{\rm ad}_Y Z)+({\rm ad}_Y X,Z)=0,$ where ${\rm
ad}_{X} Y = [X,Y]$. The perpendicular complements of the
subalgebras $a$ and $b$ with respect to this inner product take
the form,
\begin{gather}
a^{\perp} = \left\{X \in Q : \  X= X_0 + \sum_{j=1}^{\infty}
\sum_{k=1}^{2(n+2)} \frac{X_j^k} {(\lambda-\al_k)^j} \right\},
\qquad b^{\perp} = \left\{X \in Q : \  X= \sum_{j=1}^{\infty}
X_{j} \la^{j} \right\} \nonumber .
\end{gather} The natural phase space
of the ib-rMB equations, as can be seen in (\ref{ifqn}), contains
elements that belong in $a^{\perp}$ as well as terms of order
$\la^1$. Therefore we translate the space $a^{\perp}$ by
$\beta=\la X_1 \in b^{\perp}$. We note that $(\beta, [X,Y])=0$
$\forall \, X,Y \in a$, which is a necessary condition for the AKS
theorem. We consider the set,
\begin{equation}  a^{\perp} + \beta = \left\{X \in Q : \  X= X_0 + \la X_1 +
\sum_{j=1}^{\infty} \sum_{k=1}^{2(n+2)}
\frac{X_j^k}{(\lambda-\al_k)^j} \right\},
\end{equation} which includes the phase space of the ib-rMB equations, and
def\/ine a Lie--Poisson bracket on $a^{\perp} + \beta$. If $\Phi$
and $\Psi$ are functions on $a^{\perp} + \beta$ we f\/irst compute
their gradients $\nabla\Phi$, $\nabla\Psi$ in the full Lie algebra
$Q$ and then def\/ine
\begin{equation} \{\Phi,\Psi\}(X)=-\left(X,\left[ \prod_b
\nabla\Phi(X),\prod_b\nabla\Psi(X)\right] \right), \qquad \forall
\, X \in a^{\perp} + \beta  . \label{poissonbracket}\end{equation}
We note that $ \prod_{b}$ denotes projection on the Lie subalgebra
$b$. $\nabla H(X)$ denotes the gradient of a~function $H$ on $Q$
and is def\/ined as follows,
\begin{equation} (Y,\nabla H(X))=\lim_{\epsilon \rightarrow 0}
\frac{H(X+\epsilon Y)-H(X)}{\epsilon}. \label{gradients}
\end{equation}
\noindent We quote the AKS theorem and show how it can be applied
in the case of the ib-rMB equations.
\begin{theorem}
If $H$, $F$ are invariant functions on $Q^* \cong Q$, and we
denote by $H^{{\rm proj}}$, $F^{{\rm proj}}$ their projection to
the subspace $ a^{\perp} + \beta $ then $\{H^{{\rm proj}},F^{{\rm
proj}} \}=0$, and the Hamiltonian system associated with the
invariant function $H$ is given as,
\begin{equation} \frac{d}{dt} (\beta + \alpha)=-{\rm ad}^{*}_{\prod_{a}
\nabla H(\beta+\al)} (\beta + \al), \qquad \forall \, \beta+\al \in
 \beta+a^{\perp} .\end{equation}
\end{theorem}
\begin{remark}
${\rm ad}^{*}$ is negative the dual of the ${\rm ad}$ map.
\end{remark}
\begin{remark}
The def\/inition of $H$ being an invariant function is ${\rm
ad}^*_{\nabla H(x)} x=0$, $\forall \, x \in Q.$
\end{remark}
\begin{remark}
The existence of an invariant inner product allows one to identify
the ${\rm ad}$-map with ${\rm ad}^*$, because on one hand by
def\/inition $({\rm ad}_x y,z)=-(y,{\rm ad}_x^* z),$ and on the
other a simple calculation using the def\/inition of the inner
product shows that $({\rm ad}_x y,z)=-(y,{\rm ad}_x z)$.
\end{remark}
 If a set of invariant functions is given, the theorem guarantees their
commutativity in the canonical Lie--Poisson bracket and gives the
form of the Hamiltonian system associated with the invariant
function. To apply the theorem we construct a set of invariant
functions $\{\Phi_k\}_{k \in \mathbb{N}}$ via the following
operator,
\begin{equation} M_k(X)=\prod_{i=1}^{2(n+2)} (\la-\al_i)^k X. \nonumber \end{equation}
The invariant functions are def\/ined as $ \Phi_k(X)=\frac{1}{2}
(M_k(X),X). $ We compute the gradient of the $ \Phi_k(X)$. By
def\/inition, \begin{eqnarray}(Y,\nabla \Phi_k(X)) =
\lim_{\epsilon \rightarrow 0}
 \frac{\Phi_k(X+\epsilon Y)-\Phi_k(X)}{\epsilon}
  =  \left(\prod_{i=1}^{2(n+2)}(\la-\al_i)^k X,Y\right)=(Y,M_k(X)) \nonumber. \end{eqnarray}
Thus $\nabla \Phi_k(X)=M_k(X).$ According to Remarks 2 and 3, to
actually demonstrate the invariance of $\Phi_k$ we must show that
$[\nabla \Phi_k(X),X]=0$. We have that $[\nabla
\Phi_k(X),X]\!=\!\!\prod\limits_{i=1}^{2(n+2)}\!\!(\la-\al_i)^k [X,X]\!=\!0$. Therefore
the functions $\Phi_k$, $k\in \mathbb{N}$ are invariant and can be
used in the context of the AKS theorem, which reads as follows,
\begin{equation*}
\frac{d X}{dt_k} =  -\left[\prod_a \nabla \Phi_k(X),X\right]=
\left[\prod_b \nabla \Phi_k(X),X\right],
\end{equation*} since $0=[\nabla
\Phi_k(X),X]=[\prod_a \nabla \Phi_k(X),X]+[\prod_b \nabla
     \Phi_k(X),X].$
We used the subscript $k$ for the time variable $t_k$ to
distinguish between the dif\/ferent dynamical evolutions of the
systems associated with the Hamiltonian functions $\Phi_k$. For
each $k \in \mathbb{N}$ the following systems are in involution:
\begin{equation}  \frac{d X}{dt_k}= \left[\prod_b M_k(X),X\right]. \label{involution}
\end{equation} For $k=0$ and a truncated $X$ of the form (\ref{qext}) where
$h_j^i,f_j^i,e_j^i=0$ for $j \geq 2$ and $i=1,\ldots, n+2$,
\begin{equation} X=\la(h_0 \mc{H}+f_0 \mc{F}) +e_0 \mc{E}+ \sum_{i=1}^{n+2}
\frac{1}{(\lambda^2-\al_i^2)}[\lambda(h_1^i \mc{H}+f_1^i
\mc{F})+e_1^i \mc{E}],\label{x1tr} \end{equation} the Hamiltonian
system (\ref{involution}) becomes
\begin{equation} \frac{d Q^{(1)}}{dt_0}=[Q^{(0)},Q^{(1)}],\end{equation} which is the
zero curvature representation of the Lax pair equation (\ref{lpe})
for the ib-rMB equations. We have thus identif\/ied the ib-rMB
system as a member of the inf\/inite family of systems
(\ref{involution}), that commute with respect to the canonical
Lie--Poisson bracket (\ref{poissonbracket}).

\section[Extended flow]{Extended f\/low} \label{extended}

We consider a general element of the Lie algebra $Q$ of the form,
\begin{equation}  X=\la(h_0 \mc{H}+f_0 \mc{F}) +e_0 \mc{E}+
\sum_{j=1}^{\infty} \sum_{i=1}^{n+2}
\frac{1}{(\lambda^2-\al_i^2)^j}[\lambda(h_j^i \mc{H}+f_j^i
\mc{F})+e_j^i \mc{E}],\label{x1} \end{equation} or equivalently,
\begin{equation}  X=\la(h_0 \mc{H} + f_0 \mc{F}) +e_0 \mc{E} +
\sum_{j=1}^{\infty} \frac{ \sum\limits_{m=0}^{j(n+1)}
\lambda^{2m+1}(h_j^{2m+1} \mc{H}+f_j^{2m+1} \mc{F})+\la^{2m}
e_j^{2m} \mc{E}} {\prod\limits_{i=1}^{n+2} (\la^2-\al^2_{i})^j}.
\label{x2} \end{equation} We remark that $h_j^i$, $f_j^i$, $e_j^i$
in (\ref{x1}) are not the same as the ones in (\ref{x2}). The
latter ones are linear combinations of the former. However, to
avoid introducing yet another symbol we use $h_j^{2m+1}$,
$f_j^{2m+1}$, $e_j^{2m}$ in (\ref{x2}), where $j$ indicates the
order of the pole at $\la^2=\al_i^2$ and $2m+1$ or $2m$ the power
of $\la$ multiplying $\mc{H}$, $\mc{F}$ or $\mc{E}$ respectively
in the numerator.

To examine the extended f\/low associated with the ib-rMB
equations we set $k=0$ in the involutive systems
(\ref{involution}). The relevant system takes the form,
\begin{equation*}
 \frac{d X}{dt_0}= \left[\prod_b \nabla \Phi_0(X),X\right],
\end{equation*} where
$\nabla \Phi_0(X)=M_0(X)=X.$ We write X as follows:
\begin{equation*} X=\sum_{j=0}^{\infty}
\frac{P_j}{\prod\limits_{i=1}^{n+2}(\la^{2}-\al_i^2)^j},
\end{equation*}
where $P_j$ is a polynomial in $\la$ given as
\begin{equation*} P_j(\la)=\sum_{m=0}^{j(n+1)} \la^{2m+1}(h_j^{2m+1}\mc{H} +
f_j^{2m+1}\mc{F})+ \la^{2m} e_j^{2m} \mc{E}, \qquad j=1,2,\ldots,
\end{equation*} and $P_0=\la(h_0 \mc{H}+f_0 \mc{F})+e_0 \mc{E}.$ The degree of
$P_j(\la)$ is given by, $\mbox{deg}(P_j(\la))=2j(n+1)+1.$
Projecting $\nabla \Phi_0(X)$ to the subalgebra $b$ is equivalent
to keeping the polynomial part of $X$, which is $P_0$. Thus the
Hamiltonian f\/low for $k=0$ takes the form,
\begin{equation*}  \frac{d X}{dt_0}=
\sum_{j=1}^{\infty}\left[P_0,\frac{P_j}{\prod\limits_{i=1}^{n+2}(\la^2-\al_i^2)^j}\right],
\end{equation*} which unravels to
\begin{equation}  \frac{d X}{dt_0}= \sum_{j=1}^{\infty}
\left(\sum_{m=0}^{j(n+1)} \la^{2m+1}(dh_{0,j}^m \mc{H}+df_{0,j}^m
\mc{F})+ \la^{2m+2}de_{0,j}^m \mc{E} \right)\Big/
\left(\prod\limits_{i=1}^{n+2}(\la^2-\al_i^2)^j\right). \label{h1}
\end{equation} The coef\/f\/icients of the matrices $\mc{H}$, $\mc{F}$,
and $\mc{E}$ appearing in (\ref{h1}) are def\/ined as follows,
\begin{gather*}
dh_{0,j}^m=2\big(f_0 e_j^{2m}-e_0 f_j^{2m+1}\big), \qquad
df_{0,j}^m=2\big(e_0 h_j^{2m+1}-h_0 e_j^{2m}\big), \\
de_{0,j}^m=2\big(h_0 f_j^{2m+1}-f_0 h_j^{2m+1}\big) .
\end{gather*}
\noindent We observe that the expressions multiplying $\mc{H}$ and
$\mc{F}$ in (\ref{h1}) have no polynomial part since the degree of
the numerator that equals $2j(n+1)+1$  is strictly smaller that
the degree of the denominator that equals $2j(n+2)$, for $j \geq
1$. However, the expression multiplying $\mc{E}$ carries
a~polynomial term. In particular the evolution equation (\ref{h1})
can be written as
\begin{gather*} \frac{d X}{dt_0}= \sum_{j=1}^{\infty}
\left(\sum_{m=0}^{j(n+1)} \la^{2m+1}(dh_{0,j}^m \mc{H}+df_{0,j}^m
\mc{F})+  \sum_{m=1}^{j(n+1)} \la^{2m}de_{0,j}^{m-1} \mc{E} \right. \\
\left.\phantom{\frac{d X}{dt_0}=}{} +
\sum_{m=(j-1)(n+1)}^{j(n+1)} C_{m-(j-1)(n+1)}
\la^{2m}de_{0,j}^{j(n+1)} \mc{E}
\right)\Big/\left(\prod_{i=1}^{n+2}(\la^2-\al_i^2)^j\right) \\
\phantom{\frac{d X}{dt_0}=}{} + \sum_{j=2}^{\infty}
\la^{2(j-1)(n+1)}de_{0,j}^{j(n+1)}
\mc{E}\Big/\left(\prod_{i=1}^{n+2}(\la^2-\al_i^2)^{j-1}\right)+de_{0,1}^{n+1}
\mc{E}, \end{gather*} where $C_{n+1-m}=(-1)^m
\sum\limits^{n+2}_{i_{l_1} \neq i_{l_2}} \al_{i_1}^2 \cdots
\al_{i_{m+1}}^2$, $ i_{l_p} \in \{1,\ldots,n+2\}.$ On the other
hand by the def\/inition of $X$ we have that,
\begin{gather*}  \frac{d X}{dt_0}=\la(\frac{dh_0}{dt_0}
\mc{H}+\frac{df_0}{dt_0}\mc{F})+\frac{de_0}{dt_0} \mc{E} \\
 \phantom{\frac{d X}{dt_0}=}{} + \sum_{j=1}^{\infty}
\left(\la^{2m+1}(\frac{dh_j^{2m+1}}{dt_0}\mc{H}+
\frac{df_j^{2m+1}}{dt_0}\mc{F}) +
\la^{2m}\frac{de_j^{2m}}{dt_0}\mc{E}
\right)\Big/\left(\prod_{i=1}^{n+2}(\la^2-\al_i^2)^{j}\right).
\end{gather*} By
equating the dif\/ferent powers of $\la$ we disclose the system
induced by the Hamiltonian function~$\Phi_0.$ As previously noted,
for a f\/ixed $j \in \mathbb{N}$, the elements of the sets
$S_m=\{h_j^{2m+1},f_j^{2m+1},e_j^{2m}\}\!$ satisfy the same system
of equations for any $m=0,1,\ldots, j(n+1).$ Therefore it
suf\/f\/ices to consider only one such set. Without loss of
generality we choose $S_{j(n+1)}$. The evolution equations take
the form:
\begin{gather*}
\frac{dh_0}{dt_0}=\frac{df_0}{dt_0}=0, \qquad
\frac{de_0}{dt_0}= 2\big(h_0f_1^{2n+3}-f_0 h_1^{2n+3}\big), \\
\frac{dh_j^{2j(n+1)+1}}{dt_0}=2\big(f_0 e_j^{2j(n+1)}-e_0 f_j^{2j(n+1)+1}\big), \\
\frac{df_j^{2j(n+1)+1}}{dt_0}=2\big(e_0 h_j^{2j(n+1)+1}-h_0 e_j^{2j(n+1)}\big), \\
\frac{de_j^{2j(n+1)}}{dt_0}=2\big(h_0f_j^{2j(n+1)-1}-f_0
h_j^{2j(n+1)-1}\big)+2\left(\sum_{i=1}^{n+2} a_i^2\right)
\big(h_0f_j^{2j(n+1)+1}-f_0
h_j^{2j(n+1)+1}\big)\\
\phantom{\frac{de_j^{2j(n+1)}}{dt_0}=}{}+
2\big(h_0f_{j+1}^{2(j+1)(n+1)+1}-f_0 h_{j+1}^{2(j+1)(n+1)+1}\big).
\end{gather*}
To reveal the extended f\/low for the ib-rMB equations we set j=1:
\begin{gather}
\frac{dh_0}{dt_0}=\frac{df_0}{dt_0}=0, \qquad
\frac{de_0}{dt_0}=2\big(h_0f_1^{2n+3}-f_0 h_1^{2n+3}\big), \nonumber  \\
\frac{dh_1^{2n+3}}{dt_0}=2\big(f_0 e_1^{2n+2}-e_0 f_1^{2n+3}\big),
\qquad
\frac{df_1^{2n+3}}{dt_0}=2\big(e_0 h_1^{2n+3}-h_0 e_1^{2n+2}\big),  \label{exflow}\\
\frac{de_1^{2n+2}}{dt_0}=2\big(h_0f_1^{2n+1}-f_0
h_1^{2n+1}\big)+2\left(\sum_{i=1}^{n+2} a_i^2\right)
\big(h_0f_1^{2n+3}-f_0
h_1^{2n+3}\big)\nonumber\\
\phantom{\frac{de_1^{2n+2}}{dt_0}=}{}+  2\big(h_0f_2^{4n+5}-f_0
h_2^{4n+5}\big). \nonumber
\end{gather}We note that the coupling of the above system to the
evolution equations satisf\/ied by the higher order potentials
that correspond to $j \geq 2,$ is captured in the dynamical
equation for $e_1^{2n+2}.$ If $h_2^{4n+5}=f_2^{4n+5}=0,$ then the
system reduces to the dynamical equations that the $n$-soliton
potentials of the ib-rMB equations satisfy.

\section{Hamiltonian functions and Poisson brackets} \label{hf}

In this section we aim to write the extended f\/low of the ib-rMB
equations given by system (\ref{exflow}) in Section
\ref{extended}, in the canonical Poisson form,
\begin{gather}
\frac{\partial e_0}{\partial t}=\{e_0, \Phi_0\}, \qquad
\frac{\partial h_1^{2n+3}}{\partial t}=\{h_1^{2n+3}, \Phi_0\}, \label{canonical}\\
\frac{\partial f_1^{2n+3}}{\partial t}=\{f_1^{2n+3}, \Phi_0\},
\qquad \frac{\partial e_1^{2n+2}}{\partial t}=\{e_1^{2n+2},
\Phi_0\}.\nonumber
\end{gather}
The Hamiltonian functions for the systems (\ref{involution})
described in the context of the AKS theorem in Section \ref{aks}
are def\/ined as
\begin{equation} \Phi_k(X)=\frac{1}{2}(M_k(X),X)=\frac{1}{2}
\mbox{Tr}\left(\prod_{i=1}^{2(n+2)}(\la-\al_i)^k X^2 \right)_0.
\nonumber
\end{equation} We let $k=0$ and consider a general $X$ of the
form,
\begin{gather}
 X =\lambda(h_0 \mc{H}+f_0 \mc{F}) +e_0 \mc{E} \nonumber\\
\phantom{X =}{} +\sum_{j=1}^{\infty}\left(
\frac{\sum\limits_{m=0}^{j(n+1)} \la^{2m+1}(h_j^{2m+1}
\mc{H}+f_j^{2m+1} \mc{F}) + \la^{2m} e_j^{2m}
\mc{E}}{\prod\limits_{i=1}^{n+2}(\la^2-\al_i^2)^j}\right).
\nonumber \end{gather} The Hamiltonian function is found to be:
\begin{equation} \Phi_0(X)=\frac{1}{2} \mbox{Tr}\big(X^2\big)_0=e_0^2+2\big(h_0
h_1^{2n+3}+f_0 f_1^{2n+3}\big). \end{equation} \noindent We
def\/ine the following functionals, relevant to the potentials
that appear in the Hamiltonian:
\begin{alignat*}{3}
&h_0(X)=h_0 \  \mbox{coef\/f\/icient} \: \mbox{of}\:
 \la \mc{H}, \quad && h_1^{2n+3}(X)=h_1^{2n+3} \:\:
 \mbox{coef\/f\/icient}\:
\mbox{of}\:
\frac{\la^{2n+3}}{\prod\limits_{i=1}^{n+2} (\la^2-\al^2_{i})}\mc{H}, & \\
&  f_0(X)=f_0 \  \mbox{coef\/f\/icient} \:\mbox{of}\: \la \mc{F},
&&
 f_1^{2n+3}(X)=f_1^{2n+3} \:\: \mbox{coef\/f\/icient} \:\mbox{of}\:
\frac{\la^{2n+3}}{\prod\limits_{i=1}^{n+2} (\la^2-\al^2_{i})}\mc{F},& \\
 & e_0(X)=e_0 \  \mbox{coef\/f\/icient}\:  \mbox{of} \:\mc{E},
 &&
 e_1^{2n+2}(X)=e_1^{2n+2}  \:\: \mbox{coef\/f\/icient}\: \mbox{of}\:
\frac{\la^{2n+2}}{\prod\limits_{i=1}^{n+2} (\la^2-\al^2_{i})}
\mc{E}.&
\end{alignat*} Using def\/inition (\ref{gradients}) we compute the
gradients of these functionals:
\begin{alignat}{3}
& \nabla
h_0(X)=-\frac{1}{2}\frac{\la^{2n+3}}{\prod\limits_{i=1}^{n+2}
(\la^2-\al^2_{i})}\mc{H}, \qquad && \nabla h_1^{2n+3}(X)
=-\frac{1}{2}\la\mc{H} , & \nonumber\\
& \nabla
f_0(X)=-\frac{1}{2}\frac{\la^{2n+3}}{\prod\limits_{i=1}^{n+2}
(\la^2-\al^2_{i})}\mc{F}, &&  \nabla f_1^{2n+3}(X) =
-\frac{1}{2} \la \mc{F}, & \label{grad}\\
& \nabla e_0( X)=-\frac{1}{2} \mc{E}, &&  \nabla e_1^{2n+2}(X)
=-\frac{1}{2}\la^{2} \mc{E}.&\nonumber \end{alignat} For example,
to obtain $\nabla h_0(X)$ we consider the equation
\begin{equation*} (Y,\nabla h_0(X))=\lim_{\epsilon \rightarrow 0}
\frac{h_0(X+\epsilon Y)-h_0(X)}{\epsilon}, \end{equation*} which
implies that $\mbox{Tr}(Y \nabla h_0(X))_0=h_0(Y)$. Therefore
$\nabla h_0(X)$ is the element of $Q$ such that the constant term
(with respect to $\la$) of the product $Y\nabla h_0(X)$ has trace
that equals precisely~$h_0(Y).$ We note that the matrices
$\mc{H}\mc{F}$, $\mc{F}\mc{E}$, $\mc{E}\mc{H}$ are traceless
whereas $\mc{H}^2=\mc{F}^2=\mc{E}^2=-\mc{I}$. Having that in mind,
we f\/ind that $\nabla
h_0(X)=-\frac{1}{2}\frac{\la^{2n+3}}{\prod\limits_{i=1}^{n+2}
(\la^2-\al^2_{i})}\mc{H}.$

The Poisson brackets between the potentials appearing in the
Hamiltonian can be computed using def\/inition
(\ref{poissonbracket}). For instance,
\begin{gather*}  \big\{e_0,h_1^{2n+3}\big\}(X)= -\left(X,\left[\prod_{b} \nabla e_0(X),
\prod_{b} \nabla h_1^{2n+3}(X)\right]\right)\nonumber \\
\phantom{\big\{e_0,h_1^{2n+3}\big\}(X)}{} =  -\left(X,
\frac{1}{2} \la \mc{F}\right)=-\mbox{Tr}\left(\frac{1}{2} \la
X\mc{F}\right)_0 = f_1^{2n+3}. \nonumber   \end{gather*} In a
similar fashion we obtain the rest of the Poisson brackets,
\begin{gather}
\big\{e_0,h_1^{2n+3}\big\}= f_1^{2n+3}, \qquad
\big\{f_1^{2n+3},e_0 \big\}= h_1^{2n+3}, \qquad
\big\{h_1^{2n+3},f_1^{2n+3} \big\}= e_1^{2n+2}, \nonumber \\
\big\{e_1^{2n+2},h_1^{2n+3}\big\}= f_1^{2n+1} +
\left(\sum_{i=1}^{n+2} a_{i}^2 \right)f_1^{2n+3} + f_2^{4n+5},\label{pbs}  \\
\big\{f_1^{2n+3},e_1^{2n+2}\big\}= h_1^{2n+1} +
\left(\sum_{i=1}^{n+2} a_{i}^2 \right)h_1^{2n+3} + h_2^{4n+5}.
\nonumber
\end{gather}
Using the Poisson brackets (\ref{pbs}), we f\/ind that the
extended f\/low for the ib-rMB equations given in (\ref{exflow})
can be expressed as the canonical f\/low (\ref{canonical})
associated with the Hamiltonian function~$\Phi_0.$

Higher order functionals can also be def\/ined using a diagonal
formation that gradually sweeps all the potentials. In particular,
for a general $X$ of the form (\ref{x2})  we def\/ine for
$N=0,1,2,\ldots,$ the following higher order functionals (in bold
to distinguish between $\boldsymbol{h}_0$, $\boldsymbol{f}_0$,
$\boldsymbol{e}_0$ and $h_0$, $f_0$, $e_0$):
\begin{gather*}
\boldsymbol{h}_N(X)=\sum_{j=1}^{N+1}h_j^{2j(n+2)-2N-1}, \qquad
\boldsymbol{f}_N(X)=\sum_{j=1}^{N+1}f_j^{2j(n+2)-2N-1}, \\
\boldsymbol{e}_N(X)=\sum_{j=1}^{N+1}e_j^{2j(n+2)-2N-2} .
\end{gather*} We note that if the upper index of the potentials that appear
in the sums is less than zero then the potentials are set to zero.
The set $S=\{e_0,\boldsymbol{h}_N, \boldsymbol{f}_N,
\boldsymbol{e}_N:\, N=0,1,2,\ldots\}$ includes all the dynamical
quantities that enter the AKS f\/lows (\ref{involution}). The
gradients of these higher order functionals can be computed using
def\/inition (\ref{gradients}). For instance,
\begin{gather*}
\nabla \boldsymbol{h}_1(X)=-\frac{1}{2}\la^3 \mc{H} -
\left(\sum_{i=1}^{n+2}\al_i^2\right) \nabla
\boldsymbol{h}_0,\qquad \nabla
\boldsymbol{f}_1(X)=-\frac{1}{2}\la^3 \mc{F} -
\left(\sum_{i=1}^{n+2}\al_i^2\right) \nabla \boldsymbol{f}_0, \\
\nabla \boldsymbol{e}_1(X)=-\frac{1}{2}\la^4 \mc{E} -
\left(\sum_{i=1}^{n+2}\al_i^2\right) \nabla \boldsymbol{e}_0 ,
\end{gather*} where $\nabla \boldsymbol{h}_0=\nabla h_1^{2n+3}$,
$\nabla \boldsymbol{f}_0=\nabla f_1^{2n+3}$, $\nabla
\boldsymbol{e}_0=\nabla e_1^{2n+2},$ and are given in
(\ref{grad}).

Working in a similar manner as in the example for
$\{e_0,h_1^{2n+3} \},$ one can f\/ind the Poisson brackets between
the higher order functionals, i.e. $\{e_0,\boldsymbol{h}_1\}(X)=
\boldsymbol{f}_1(X).$

\section{Summary}

In this paper we have considered an integrable system of reduced
Maxwell--Bloch equations, that is inhomogeneously broadened. We
show that the relevant B\"acklund transformation preserves the
reality of the $n$-soliton potentials $\forall \, n \in
\mathbb{N},$ and establish their pole structure with respect to
the broadening parameter. We obtain a representation of the
relevant phase space in the spectral parameter $\la$, which is
then embedded in a prolonged loop algebra. The equations
satisf\/ied by the $n$-soliton potentials are associated to an
inf\/inite family of higher order Hamiltonian involutive systems.
We present the Hamiltonian functions of the higher order f\/lows
and the Poisson brackets between the extended potentials.

\subsection*{Acknowledgements}

The author would like to thank P. Shipman for useful discussions
and the Cyprus Research Promotion Foundation for support through
the grant CRPF0504/03.

\LastPageEnding

\end{document}